\documentclass[aip,reprint,amsmath,amssymb]{revtex4-1}

\usepackage{graphicx}
\usepackage{dcolumn}
\usepackage{float}
\usepackage{bm}
\usepackage{amssymb}
\usepackage{natbib}

\begin{document}


\title{Polymer Translocation Dynamics in the Quasi-Static Limit}

\author{James M. Polson and Anthony C. M. McCaffrey}
\affiliation{%
Department of Physics, University of Prince Edward Island, 550 University Ave.,
Charlottetown, Prince Edward Island, C1A 4P3, Canada
}%

\date{\today}

\begin{abstract}
Monte Carlo (MC) simulations are used to study the dynamics of polymer translocation
through a nanopore in the limit where the translocation rate is sufficiently
slow that the polymer maintains a state of conformational quasi-equilibrium.
The system is modeled as a flexible hard-sphere chain that translocates through 
a cylindrical hole in a hard flat wall. In some calculations, the nanopore is
connected at one end to a spherical cavity.  Translocation times are measured 
directly using MC dynamics simulations.  For sufficiently narrow pores, translocation 
is sufficiently slow that the mean translocation time scales with polymer length $N$ 
according to $\langle \tau\rangle \propto (N-N_{\rm p})^2$, where $N_{\rm p}$ is
the average number of monomers in the nanopore; this scaling is an indication 
of a quasi-static regime in which polymer-nanopore friction dominates. 
We use a multiple-histogram method to calculate the variation of the 
free energy with $Q$, a coordinate used to quantify the degree of translocation. 
The free energy functions are used with the Fokker-Planck formalism to calculate
translocation time distributions in the quasi-static regime. These calculations 
also require a friction coefficient, characterized by a quantity $N_{\rm eff}$,
the effective number of monomers whose dynamics are affected by the confinement
of the nanopore. This was determined by fixing the mean of the theoretical 
distribution to that of the distribution obtained from MC dynamics simulations. 
The theoretical distributions are in excellent quantitative agreement with
the distributions obtained directly by the MC dynamics simulations for physically
meaningful values of $N_{\rm eff}$.  The free energy functions for narrow-pore systems 
exhibit oscillations with an amplitude that is sensitive to the nanopore length. 
Generally, larger oscillation amplitudes correspond to longer translocation times. 
\end{abstract}

\maketitle

\section{Introduction}
\label{sec:intro}
The translocation of a polymer through a narrow passage from one space to
another is a fundamental process with numerous applications.\cite{Muthukumar_book}
It is a key part of various biological phenomena, including viral genome transfer,
RNA transport through the nuclear pore complex, protein transport into mitochondria, 
and genome packing in bacteriophages.\cite{Alberts_book,Lodish_book}
The pioneering work of Kasianowicz {\it et al.}  first demonstrated the feasibility of
monitoring translocation of single molecules by measurement of the ionic
current passing through nanometre-sized pores.\cite{Kasianowicz_1996,Bezrukov_1996}
In this method, the potential difference that drives the ionic current
also drives a charged macromolecule through
a nanopore, and the translocation results in a measurable reduction in the current.
Substantial effort has been devoted to further develop this experimental
technique using both biological\cite{Akeson_1999,Meller_2000,Meller_2003,Wang_2004,%
Butler_2006,Wong_2008,Wong_2010} and synthetic nanopores.\cite{Li_2003,Chen_2004,%
Fologea_2005a,Fologea_2005b,Storm_2005,Storm_2005b,Purnell_2008}
This work has largely been motivated by the desire to
develop a fast and inexpensive method of sequencing DNA.%
\cite{Branton_2008_abbr,Venkatesan_2011,Wanunu_2012} 
The experiments have revealed a complex dependence of translocation behaviour on 
a variety of factors, including electric field strength,\cite{Storm_2005} pH,\cite{Wong_2010}
and salt concentration.\cite{Kowalczyk_2012} An understanding of this growing
body of experimental results is currently incomplete.\cite{Muthukumar_book}

Numerous theoretical and computer simulation studies of polymer translocation
have been reported in recent years. 
Much of this work has been summarized and evaluated in a recent review.\cite{Milchev_2011} 
One notable theoretical approach, pioneered by Sung and Park\cite{Sung_1996} and
Muthukumar\cite{Muthukumar_1999} employs the Fokker-Planck (FP) formalism. 
Here, the dynamics are governed by the free energy,
$F(s)$, which is a function of a single coordinate, $s$, typically defined to be
the number of monomers on one side the nanopore. The time-dependent probability
distribution for this coordinate is obtained by solving a 1-D Smoluchowski
equation, for which the free energy function is required input. This solution can be
used to calculate translocation time distributions and determine scaling relations 
for mean translocation times.  The predicted scaling relations 
are quantitatively consistent with results from experiments of DNA 
translocation through $\alpha$-hemolysin.\cite{Muthukumar_book,Muthukumar_1999}
The FP formalism has been used to study translocation to various levels
of approximation and for different confinement geometries.\cite{Muthukumar_2001,%
Muthukumar_2003,Slonkina_2003,Kong_2004,Kejian_2006,Wong_2008,Mohan_2010,Yang_2012,%
Rasmussen_2012,Mirigian_2012,Qian_2012}
A detailed review of the FP formalism and its application to polymer translocation
is provided in the recent book by Muthukumar.\cite{Muthukumar_book}

The veracity of the free energy functions clearly determines the accuracy
of the predictions for translocation dynamics in the FP approach.
Most FP theoretical studies employ analytical approximations of $F(s)$%
\cite{Sung_1996,Muthukumar_1999,Slonkina_2003,Muthukumar_2003,Slonkina_2003,Kong_2004,
Kejian_2006,Gauthier_2008a,Gauthier_2008b,Wong_2008,Mohan_2010,deHaan_2011a,Yang_2012} 
In other studies, the free energies have been calculated numerically for 
specific molecular models that can also be employed in dynamics simulations.
For example, exact enumeration of polymer configurations
has been used to calculate a free energy landscape in order to elucidate
the effects of polymer-nanopore attractions on translocation times for 
a lattice model system.\cite{Sun_2009,Chen_2009}
Off-lattice models have been studied using specialized Monte Carlo simulation methods. 
One recent example is the incremental gauge cell (IGC) method,\cite{Rasmussen_2011}
which was used to calculate free energy profiles for a flexible Lennard-Jones chain
translocating out of a spherically confined space with adsorbing walls.\cite{Rasmussen_2012}
Another efficient MC method appropriate 
to calculate translocation free energies is the self-consistent multiple histogram (SCMH)
technique.\cite{Frenkel} Recently, we used the SCMH method to calculate free energy functions
for flexible hard-sphere chains translocating through a cylindrical pore in a hard
barrier.\cite{Fatehi_2009,Polson_2013} 
Generally, the free energy curves follow trends that conform well to the 
predictions of a modified version of the standard analytical model described 
in Ref.~\onlinecite{Muthukumar_book}. 
However, one unexpected feature observed was an oscillation in the free energy
with an amplitude that is very sensitive to the nanopore dimensions.\cite{Polson_2013} 
While its relevance to the experimental context is unclear, knowledge
of such effects may be helpful for the interpretation of results from
dynamics simulations and could be relevant to elucidating the sensitivity of the translocation 
behaviour on seemingly minor differences in the model that have been observed in some 
simulation studies.\cite{Lehtola_2010a,deHaan_2010,Panja_2010}

The validity of the FP approach to polymer translocation
is founded on the assumption that the portions of the polymer outside the
pore remain in conformational equilibrium during the entire translocation process.
This is expected to be the case in the limit of sufficiently strong pore friction
for polymers of finite length. In this case the translocation time scales
as $N^2$ for unbiased translocation\cite{Muthukumar_1999,deHaan_2011a} and as 
$N/f$ for forced translocation,\cite{Muthukumar_1999} where $f$ is the driving force 
magnitude.  However, it is well known that this assumption must break down for 
sufficiently long polymers.\cite{Chuang_2001,Kantor_2004}
Computer simulation studies have generally yielded scaling exponents 
inconsistent with the values for the quasi-static regime with strong pore friction.%
\cite{Chuang_2001,Milchev_2004,Kantor_2004,Tsuchiya_2007,Luo_2006a,Wolterink_2006,%
Guillouzic_2006,Dubbeldam_2007a,Dubbeldam_2007b,Huopaniemi_2007,Panja_2007,Panja_2008a,%
Wei_2007,Luo_2008a,Vocks_2008,Lehtola_2008,Lehtola_2009,Luo_2009,%
Gauthier_2009,Bhattacharya_2009,Kapahnke_2010,deHaan_2010,Dubbeldam_2011,deHaan_2012b,%
Ikonen_2012a} The values of the exponents vary widely and appear to be sensitive to 
the details of the model, simulation method, and the length of the polymer.
In addition, nonequilibrium conformational behaviour              
has been observed directly.\cite{Luo_2006b,Gauthier_2009,%
Bhattacharya_2009,Luo_2010,Bhattacharya_2010,Dubbeldam_2012,Feng_2012}
Various theoretical models have been developed to account for
non-equilibrium effects and generally predict scaling exponent
values greater than those for the quasi-equilibrium case,
both for nondriven\cite{Panja_2007,Dubbeldam_2007a,deHaan_2012b} 
and driven\cite{Dubbeldam_2007b,Vocks_2008,Sakaue_2007,Sakaue_2010,Saito_2011,%
Rowghanian_2011,Dubbeldam_2012,Ikonen_2012a,Ikonen_2012b,Ikonen_2012c} 
translocation. Clearly, the FP formalism is an inappropriate approach
for the conditions considered in most of these simulation studies.
However, simulation models can easily be adjusted to either
reduce the relaxation time of the chain or increase the 
nanopore friction, either of which should eventually lead to quasi-static conditions.
This is illustrated by a recent simulation study of unforced translocation 
that followed the former approach and measured a scaling exponent 
value consistent with FP predictions for the quasi-static regime 
at sufficiently low viscosity where the relaxation time of the
chain is very short.\cite{deHaan_2012b} 

The purpose of this study is to examine the accuracy with which the
FP formalism describes the translocation dynamics in the quasi-static regime. 
The system is modelled as freely-jointed hard-sphere chain that translocates through a 
cylindrical nanopore in a hard barrier. We employ free energy functions 
that were calculated using the SCMH method in our recent study.\cite{Polson_2013} 
The  calculated translocation time distributions are compared to those
measured directly from MC dynamics calculations for the same model.
Calculations were carried out for systems in the quasi-static regime with friction
dominated by the nanopore. The regime boundary is determined by measurement 
of the scaling exponent in MC dynamics simulations. The friction coefficient 
used in the FP calculations is characterized by a parameter $N_{\rm eff}$,
the effective number of monomers whose dynamics are affected by nanopore confinement.
We focus mainly on a system with a planar barrier and semi-infinite {\it cis} and 
{\it trans} spaces.  We also present some results for the cases where the pore is connected
to a spherical cavity at one end; comparable systems have been the focus of
a number of other theoretical and computer simulation studies.\cite{Muthukumar_2003,%
Muthukumar_2001,Kong_2004,Cacciuto_2006,Mohan_2010,Zhang_2012,Yang_2012}
We investigate the effects on the translocation dynamics of varying the nanopore dimensions,
the initial position of the polymer, and polymer length. 
Generally, the theoretical distributions are in excellent quantitative agreement
with those measured from MC dynamics simulations for physically meaningful values
of $N_{\rm eff}$.


The remainder of this article is organized as follows.  
In Section \ref{sec:model}, the model is described.
In Section~\ref{sec:theory}, we present a brief overview of the relevant
results of the FP methodology and describe how the friction
coefficient employed in the theory is calculated.
In Section~\ref{sec:methods}, we describe the MC algorithm used 
in the dynamics simulations and the details of the FP calculations. 
In Section~\ref{sec:results}, the simulation results are presented, and 
the significance of the results is discussed in Section~\ref{sec:discussion}.

\section{Model}
\label{sec:model}

The polymer is modeled as a flexible chain of $N$ hard spheres, each with a diameter
of $\sigma$. The pair potential for nonbonded monomers is thus
\begin{equation}
u_{\rm{nb}} (r)=
\begin{cases}
 \infty , \hspace{8pt} r<\sigma \cr 0 , \hspace{8pt} r>\sigma,
\end{cases}
\label{eq:non-bonded}
\end{equation}
where $r$ is the distance between the centres of the monomers. Pairs of bonded monomers
interact with a potential
\begin{equation}
u_{\rm{b}}(r)=
\begin{cases}
\hspace{1mm} 0 , \hspace{14pt} 0.9\sigma<r<1.1\sigma \cr \infty , 
\hspace{8pt} \rm{otherwise}.
\end{cases}
\label{eq:bonded}
\end{equation}
Thus, the bond length can fluctuate slightly about its average value, which is equal to the
monomer diameter.

\begin{figure}[htbp]
\centering
\includegraphics[width=0.35\textwidth]{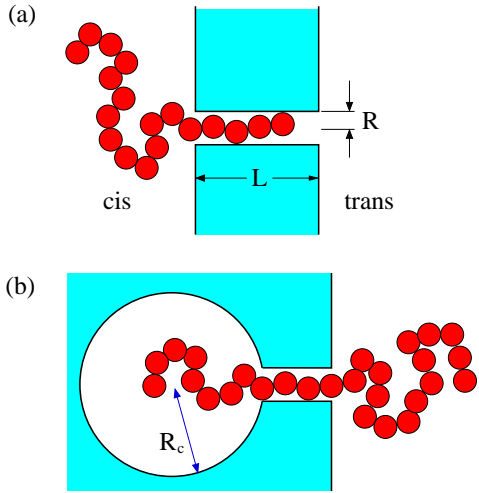}
\caption{Illustration of the two confinement geometries employed in this study.
Most of the calculations in this study used the system shown in (a).
}
\label{fig:geometry}
\end{figure}

The polymer undergoes translocation through a narrow pore from one region to another.
The nanopore is modeled as a cylindrical hole of length $L$  and radius $R$ in a hard barrier. 
The regions on the two sides of the nanopore are labeled {\it cis} and {\it trans}. Two
different cases are considered in this study. In the first case, the {\it cis} and {\it trans}
sides are semi-infinite spaces, and the barrier that separates them is bounded by two
flat surfaces. Monomers cannot overlap with the barrier, and thus the minimum distance 
between the centre of a monomer and the nearest point on either the
barrier wall surface or the nanopore surface is $\sigma/2$. This confinement geometry is
illustrated in Fig.~\ref{fig:geometry}(a). In the second case, translocation occurs
between a spherical cavity of radius $R_{\rm c}$ on the {\it cis} side to a semi-infinite
space on the {\it trans} side, as shown in Fig.~\ref{fig:geometry}(b). 
Most of the calculations in this study employed the first variant of the model.

The degree to which the polymer has translocated across the nanopore is quantified
using a translocation coordinate, $Q$, which is defined:
\begin{equation}
Q = \frac{1}{N} \sum_{i=1}^N Q_i,
\label{eq:Qdef}
\end{equation}
where the translocation coordinate for the $i$th monomer is given by
\begin{equation}
Q_i  =
\begin{cases}
0, z_i < 0 \cr
z_i/L, 0 < z_i < L \cr
1, z_i > L
\label{eq:Qidef}
\end{cases}
\end{equation}
where $z_i$ is the $z$-coordinate for the $i$th monomer. Thus, $Q=0$ when all monomers
are on the $cis$ side of the pore, $Q=1$ when the polymer is completely on the $trans$
side, and $0<Q<1$ if there are any monomers inside the nanopore.
Clearly, $Q$ is a continuous quantity, which makes it a more convenient choice
for a translocation coordinate here than the quantity $s$ (the number of monomers 
on the {\it trans} side of the pore) used in most translocation studies.

\section{Theoretical Background}
\label{sec:theory}

In the quasi-static limit, the time-dependent translocation probability distribution,
${\cal P}(Q,t)$, is governed by the Fokker-Planck equation.
When the pore friction is dominant contribution to the total polymer friction
the form of the equation is\cite{Muthukumar_book}
\begin{eqnarray}
\frac{\partial{\cal P}}{\partial t} & = & D\left[
\frac{\partial}{\partial Q} \left(\frac{1}{k_{\rm B}T} \frac{\partial F}{\partial Q}
{\cal P}(Q,t)\right) + \frac{\partial^2{\cal P}(Q,t)}{\partial Q^2} \right],
\label{eq:FP}
\end{eqnarray}
where $k_{\rm B}$ is Boltzmann's constant, $T$ is temperature,
$F(Q)$ is the equilibrium translocation free energy function and
$D$ is an effective diffusion coefficient. 
Consider a domain bounded by $Q=Q_{\rm a}$ and $Q=Q_{\rm b}$.
The first passage time, $\tau$, for a polymer located at $Q_0 \in [Q_{\rm a},Q_{\rm b}]$
at time $t=0$ to reach either boundary has a probability distribution given 
by\cite{Muthukumar_book}
\begin{eqnarray}
g(\tau;Q_0) = -\frac{d}{d\tau} \int_{Q_a}^{Q_b} dQ \,p(Q,\tau;Q_0,0),
\label{eq:gQ0tau}
\end{eqnarray}
where $p(Q,t;Q',t')$ is the conditional probability of the polymer reaching 
a value $Q$ at time $t$ given that it had a value of $Q'$ at an earlier time $t'$.
For a Markov process, $p(Q,t;Q',t')$ also satisfies Eq.~(\ref{eq:FP}). 
In the case of absorbing boundary conditions the mean first passage time is 
given by:\cite{Muthukumar_book}
\begin{widetext}
\begin{eqnarray}
\langle\tau\rangle = \frac{\left( \int_{Q_0}^{Q_{\rm b}} \frac{dy}{\psi(y)} \right)
\int_{Q_{\rm a}}^{Q_0} \frac{dy'}{\psi(y')} H(y') -  \left( \int_{Q_{\rm a}}^{Q_0} 
\frac{dy}{\psi(y)} \right) \int_{Q_0}^{Q_{\rm b}} \frac{dy'}{\psi(y')} H(y') } 
{ \int_{Q_{\rm a}}^{Q_{\rm b}}\frac{dy}{\psi(y)} },
\label{eq:tauQ0}
\end{eqnarray}
\end{widetext}
where
\begin{eqnarray}
\psi(Q) = \exp\left(-[F(Q)-F(Q_{\rm a})]/k_{\rm B}T\right).
\label{eq:psiQ}
\end{eqnarray}
and
\begin{eqnarray}
H(Q) & = & - D^{-1} \int_{Q_{\rm a}}^Q dQ' \psi(Q')
\label{eq:HQ}
\end{eqnarray}
Finally, the probability a polymer reaches $Q=Q_{\rm b}$ without ever
reaching $Q_{\rm a}$ is given by:\cite{Muthukumar_book}
\begin{eqnarray}
\pi_{\rm b}(Q_0) = \frac{\Psi(Q_{\rm a},Q_0)}{\Psi(Q_{\rm a},Q_{\rm b})}
\label{eq:pib}
\end{eqnarray}
where 
\begin{eqnarray}
\Psi(x,y) \equiv \int_x^y\frac{dz}{\psi(z)}
\end{eqnarray}
Note that this probability depends only on the free energy function and not on the
diffusion coefficient.

In order to calculate $g(\tau;Q_0)$ and $\langle\tau\rangle$, the free energy function, $F(Q)$,
and the diffusion coefficient $D$ must be supplied. In this study, $F(Q)$ is calculated using
the SCMH method, as described in Ref.~\onlinecite{Polson_2013}.
The diffusion coefficient $D$ is closely related to the diffusion coefficient of
monomers confined by the nanopore. In this study, we employ a MC method 
to study the translocation dynamics of the polymer; the details of the method
will be described below in Section~\ref{subsec:theory-dynamics}. The diffusion rate for
monomers inside the nanopore is determined by nature of the MC trial moves
for the individual monomers. Unfortunately, quantitatively accurate values of 
$D$ cannot easily be calculated analytically from knowledge of the
details of the model and method alone. Consequently, the diffusion coefficient is 
essentially treated as a free parameter. However, it is possible to determine the 
scaling of $D$ with polymer length and (approximately) with nanopore length 
in the limit of strong pore friction.

Consider a polymer that moves inside an infinitely long cylindrical tube aligned along
the $z$ axis. The polymer diffuses along $z$ such that its mean square displacement 
satisfies
\begin{eqnarray}
\langle \Delta z^2\rangle = 2 D_z t = \frac{2k_{\rm B}T}{\gamma_z} t
\label{eq:Dz2gz}
\end{eqnarray}
where $D_z=k_{\rm B}T/\gamma_{\rm z} $ is the diffusion coefficient and $\gamma_z$ is 
the friction coefficient associated with the polymer inside the tube.  
In this study, we consider the model polymer described in Section~\ref{sec:model} 
that moves via random displacements of randomly chosen monomers, one at
a time.  In this case, the friction coefficient in Eq.~(\ref{eq:Dz2gz}) scales as
\begin{eqnarray}
\gamma_z = \gamma_0 (N-1),
\label{eq:gz0_inf}
\end{eqnarray}
where $N$ is the number of monomers. In addition, $\gamma_0$ is a constant that depends
on the radius of the tube and the probability distribution of trial monomer displacements
employed in the calculations. This scaling relation is confirmed through results
from simulations.
Now consider a polymer undergoing translocation through a cylindrical nanopore
of finite length, $L$. For a sufficiently narrow pore, the pore friction is
expected to be the dominant contribution to the overall friction of the polymer. 
In this case, the friction coefficient governing 
the motion along $z$ for the portion of the polymer inside the pore is expected to 
obey a relation similar Eq.~(\ref{eq:gz0_inf}),
\begin{eqnarray}
\gamma_z = \gamma_0 (N_{\rm eff}-1),
\label{eq:gz0}
\end{eqnarray}
where $N_{\rm eff}$ is the effective number of monomers whose dynamics are strongly 
effected by the presence of the nanopore. 
To estimate $N_{\rm eff}$, first note that the range over which monomers are subject to
some degree of confinement is $L+\sigma$. For a narrow nanopore, the polymer is approximately
linear inside the pore and the number of bonds spanning
the confinement region is $n_{\rm bond} = (L+\sigma)/\sigma$; thus, the number of
monomers spanning this region can be estimated as $N_{\rm eff} \approx n_{\rm bond}+1$, or
\begin{eqnarray}
N_{\rm eff} \approx L/\sigma + 2.
\label{eq:Npapprox}
\end{eqnarray}
In practice, $N_{\rm eff}$ is a free parameter, albeit one that should come close
to satisfying the approximation in Eq.~(\ref{eq:Npapprox}). 

In the absence of a free energy gradient, and for sufficiently strong pore friction,
the mean square displacement for monomers
inside the nanopore is expected to obey Eq.~(\ref{eq:Dz2gz}). 
From the definition of $Q$ in Eqs.~(\ref{eq:Qdef}) and (\ref{eq:Qidef}) it follows
that a pore monomer displacement of $\Delta z$ corresponds to a translocation
coordinate change of $\Delta Q = \Delta z/(N\sigma)$, on average. Consequently,
$\left\langle \Delta Q^2\right\rangle = 2 D t$, where 
\begin{eqnarray}
D = \frac{k_{\rm B}T}{\gamma},
\label{eq:Dg}
\end{eqnarray}
and where $\gamma = N^2\gamma_z$.
$D$ is the diffusion coefficient that appears in Eq.~(\ref{eq:FP}), and $\gamma$
is the corresponding friction coefficient. It follows that
\begin{eqnarray}
\gamma = N^2\gamma_0(N_{\rm eff}-1).
\label{eq:gN2Np}
\end{eqnarray}
The quantity $N_{\rm eff}$ is mainly determined by the nanopore length and can be  approximated
using Eq.~(\ref{eq:Npapprox}). The quantity $\gamma_0$ is governed by the specifics
of the MC dynamics algorithm. In practice, it can be determined from MC measurements 
of the time-dependence of $\langle \Delta z^2\rangle$ for polymers of different lengths 
confined to an infinite cylindrical tube. 

\section{Methods}
\label{sec:methods}

\subsection{Free Energy Calculations}
\label{subsec:theory-free-energy}

Monte Carlo simulations employing the Metropolis algorithm and the self-consistent 
multiple histogram (SCMH) method\cite{Frenkel} were used to calculate the free 
energy functions for the polymer-nanopore model described in Section~\ref{sec:model}. 
The results of these calculations were presented in a recent article,\cite{Polson_2013}
to which the reader should refer for a detailed description of the methodology.

\subsection{MC Dynamics Simulations}
\label{subsec:theory-dynamics}

Monte Carlo dynamics simulations were used to study the translocation dynamics for
the system. Polymer motion was generated through random monomer displacement, in which
the coordinates of a randomly chosen monomer were displaced by an amount
$\Delta r_{\lambda}$ for $\lambda=x,y,z$. Each coordinate displacement was randomly 
chosen from a uniform distribution $[-\Delta_{\rm max},\Delta_{\rm max}]$. Unless 
otherwise stated, $\Delta_{\rm max}=0.14\sigma$. A trial moves was rejected if 
it lead to overlap with another nonbonded monomer, violation of the bonding
constraint, or overlap with the nanopore or barrier wall.  The polymer
coordinates were chosen to correspond a desired initial coordinate, $Q_0$, and
the system was equilibrated for fixed $Q_0$ for typically $10^6$ MC cycles.
Two different first passage times
were measured: (1) $\tau_1$, the first time that all monomers had completely
emptied from either the {\it cis} or {\it trans} domain, and (2) $\tau$, the first
time that all monomers had completely emptied the nanopore and were all in either
the {\it cis} or {\it trans} region. We define $\tau_2$ to be the time taken by the polymer to
empty the nanopore, and thus, $\tau = \tau_1 + \tau_2$.
We simulated typically 500 -- 2000 translocation events to calculate mean translocation
times and $10^4-10^5$ events in cases where translocation time distributions were desired.

\subsection{Theoretical Calculations}
\label{subsec:theory-calculations}

MC dynamics simulations of a hard-sphere chain polymer in an infinitely long cylinder
were used to measure $\langle \Delta z^2\rangle$ as a function of time for the
centre of mass of a polymer for several lengths. 
A linear fit to each of these functions was used to extract 
the friction coefficient $\gamma_z$ using Eq.~(\ref{eq:Dz2gz}). Fitting $\gamma_z$ vs
polymer length using Eq.~(\ref{eq:gz0_inf}) yielded the constant $\gamma_0$ for a tube of a 
specified radius.

Equations~(\ref{eq:tauQ0}), (\ref{eq:psiQ}) and (\ref{eq:HQ}) were used to calculate 
the mean first passage time for the translocation events. The integrals were calculated
numerically using the trapezoid rule. The friction coefficient $\gamma$ was calculated using
Eq.~(\ref{eq:gN2Np}), and the corresponding diffusion coefficient, $D$, was then calculated
using Eq.~(\ref{eq:Dg}). The value of the effective number of monomers in the nanopore, 
$N_{\rm eff}$, was chosen so that the calculated $\langle\tau_1\rangle$ was equal to the
value measured directly in the MC dynamics simulations.  After determining $D$, 
Eq.~(\ref{eq:FP}) was solved numerically using a simple finite-difference method
with the boundary condition $Q=Q_0$ at $t=0$ to yield the function ${\cal P}(Q,t)$.
%
%
The distribution of first passage
times, $g(\tau;Q_0)$ was calculated by solving Eq.~(\ref{eq:gQ0tau}); the integral
was first solved numerically using the trapezoid rule, and the derivative was calculated
using the finite-difference approximation.

In the results presented below, lengths are measured in units of $\sigma$, the
monomer diameter. In addition, time is measured in MC cycles, where 1 MC cycle corresponds
to one attempted move per monomer, on average.

\section{Results}
\label{sec:results}

The focus of this study is the elucidation of polymer translocation dynamics in
the quasi-static regime. In this regime, the FP formalism described in the previous
section is valid and can be used to calculate first passage times associated with translocation.
For a given model system, it is not clear {\it a priori} which parameter values 
correspond to this regime. One effective means to delineate the regime boundaries 
is to measure the scaling exponent $\alpha$, defined by the power law scaling relation:
\begin{eqnarray}
\langle\tau\rangle \propto N^\alpha.
\label{eq:tauN}
\end{eqnarray}
In the limit of sufficiently strong pore friction, where the system is in the
quasi-static regime, it is expected that $\alpha = 2$ for unforced translocation. 
For a nanopore of finite length $L$, the most appropriate scaling relation is slightly 
different.  For a narrow nanopore where the polymer has a linear 
configuration inside the nanopore, the appropriate relation is:
\begin{eqnarray}
\langle\tau\rangle \propto (N-N_{\rm p})^\alpha,
\label{eq:tauN2}
\end{eqnarray}
where 
\begin{eqnarray}
N_{\rm p} \equiv L/\sigma + 1
\end{eqnarray}
is approximately the number of monomers that lie inside the nanopore. The rationale
for this modification is that most of the time for translocation passes during the
phase in which the nanopore is filled, and monomers lie outside in both the
{\it cis} and {\it trans} regions, while the time elapsed during the pore-emptying
stage is very small, by comparison. Thus, translocation is almost complete by the time
either the {\it cis} or {\it trans} region first empties at time $\tau_1$, and
$\tau = \tau_1 + \tau_2 \approx \tau_1$, since $\tau_2 \ll \tau_1$.
Justification for this claim will be presented below.
From this perspective, the effective length of the polymer is $N-N_{\rm p}$. 
Of course, in the limit that $N \gg N_p$, Eq.~(\ref{eq:tauN2}) reduces to 
Eq.~(\ref{eq:tauN}). 

\begin{figure}[!ht]
\includegraphics[width=0.40\textwidth]{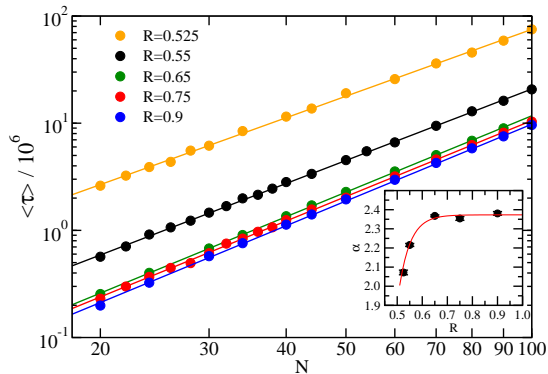}
\caption{Mean translocation time, $\langle\tau\rangle$, vs polymer length, $N$, for a pore
of length $L=0.1$ for several different pore radii. The inset shows the scaling
exponent, $\alpha$, vs pore radius $R$ obtained from fits to the data. The solid
red curve is a guide for the eye.}
\label{fig:Nscale.L=0.1}
\end{figure}

It is instructive to first compare the scaling results in this MC dynamics study with those
from the Langevin dynamics study of Ref.~\onlinecite{deHaan_2010}, which examined
polymer translocation through a hole in a flat two dimensional barrier.
Because of the variability of bond lengths permitted by Eq.~(\ref{eq:bonded}), 
we choose a small, but finite, barrier width of $L=0.1$.
This precludes the unphysical possibility that
two bonded monomers are bisected by the wall away from the pore. This barrier width 
should be small enough to approximate a 2-D barrier.  Figure~\ref{fig:Nscale.L=0.1} 
shows the mean translocation time $\langle\tau\rangle$ vs polymer length $N$ for 
several different pore radii.  In these simulations, the polymer was initially placed 
midway in the nanopore such that $Q_0=0.5$.  Over the range of $N$ considered 
($N=20$--$100$) $\langle\tau\rangle$ satisfies the power relation
of Eq.~(\ref{eq:tauN}).  The solid curves show the best fits to the simulation data.
The inset of the figure shows the corresponding values of the scaling exponents, $\alpha$, 
as a function of pore radius, $R$. The exponent has a value
of $\alpha\approx 2.4$ in the range $R=0.65$--0.9, but decreases rapidly with
decreasing $R$ for $R<0.65$ and reaches $\alpha=2.07\pm 0.01$ for $R=0.525$.
These data are in general agreement with the results of Ref.~\onlinecite{deHaan_2010}. 
Note that this scaling behaviour cannot be extrapolated to very large $N$.
As $N$ increases, the scaling is expected to change in a manner such that 
$\langle \tau\rangle$ increases monotonically with decreasing $R$.

\begin{figure}[!ht]
\includegraphics[width=0.40\textwidth]{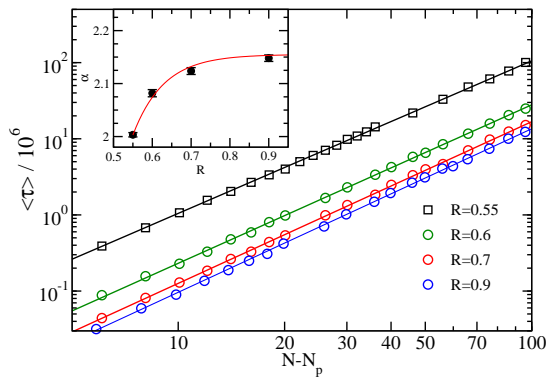}
\caption{Mean translocation time, $\langle\tau\rangle$, vs $N-N_{\rm p}$, for a pore
of length $L=2.917$ for several different pore radii. The inset shows the scaling
exponent, $\alpha$, vs pore radius $R$ obtained from fits to the data. The solid
red curve is a guide for the eye. }
\label{fig:tau.Nscale.L=2.917}
\end{figure}

Figure~\ref{fig:tau.Nscale.L=2.917} shows similar scaling data for a nanopore of
length $L=2.917$.  As in the case described above, decreasing $R$ results
in an increase in $\langle\tau\rangle$. Consequently, the polymer
can more easily maintain a state of conformational equilibrium as translocation proceeds,
which leads to the observed decrease in the scaling exponent with $R$.
The scaling exponent reaches a value of $\alpha=2.003\pm 0.004$ for a nanopore 
radius of $R=0.55$, which corresponds to the case of quasi-static dynamics and 
strong pore friction.
%
%

\begin{figure}[!ht]
\includegraphics[width=0.40\textwidth]{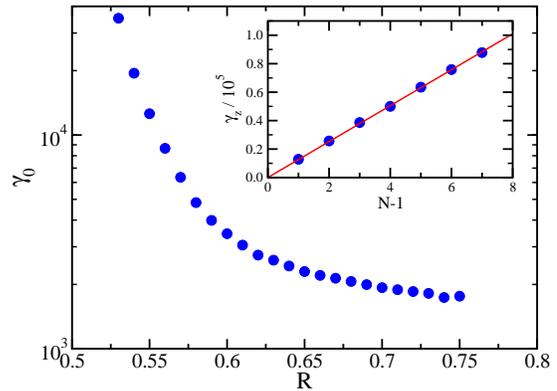}
\caption{Friction parameter, $\gamma_0$, vs nanopore radius, $R$, for a polymer
in an infinitely long cylindrical tube. The inset shows the friction coefficient
$\gamma_z$, vs polymer length, $N-1$, for
a polymer in an infinite cylindrical tube of radius $R=0.55$. The red curve
is a fit to the linear function of Eq.~(\ref{eq:gN2Np})
}
\label{fig:gamma.R}
\end{figure}

One reason that decreasing the pore radius increases $\langle\tau\rangle$ 
is related to the variation of the friction parameter $\gamma_0$ with $R$.
As $R$ decreases, the acceptance probability of trial displacements of monomers
inside the pore decreases because of an increasing likelihood that the move
leads to overlap with the nanopore wall.  This reduces the
mobility for these monomers and therefore increases in $\gamma_0$.
This trend is illustrated in Fig.~\ref{fig:gamma.R}.
From Eq.~(\ref{eq:gN2Np}), it follows that the friction coefficient for the translocating
polymer, $\gamma$, should also increase.  
The inset of Fig.~\ref{fig:gamma.R} shows the variation of the friction coefficient 
for a polymer in an infinite cylindrical tube, $\gamma_z$, as a function of $N-1$ 
for $R=0.55$, where each $\gamma_z$ value was determined by measuring 
$\langle \Delta z^2\rangle$ as a function of time and using Eq.~(\ref{eq:Dz2gz}). 
Fitting the data using Eq.~(\ref{eq:gz0_inf}) yields the value of $\gamma_0$. 
These results confirm the scaling relation of Eq.~(\ref{eq:gz0_inf}).
[Note that the coupling between $\gamma_0$ and $R$ is a consequence
of the MC dynamics algorithm used.  An alternative choice of trial moves 
for monomers in the pore, e.g. smaller displacements in the $x$-$y$ 
direction compared to those along $z$, would allow decoupling of friction and pore 
radius.]

\begin{figure}[!ht]
\includegraphics[width=0.40\textwidth]{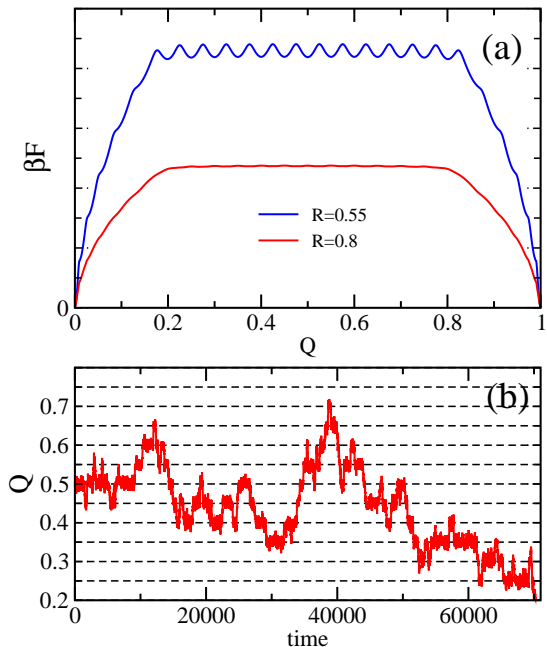}
\caption{(a) Translocation free energy profiles for a polymer of length $N=20$ in a nanopore
of length $L=6$ for two different pore radii. (b) $Q$, vs time for an escape event 
for $R=0.55$. The horizontal dashed lines mark the values of $Q$ corresponding to 
local free energy minima.}
\label{fig:F.N20.2R}
\end{figure}

A second factor contributing to slower translocation as $R$ decreases
can be understood from the effect of decreasing pore radius on the free energy
profiles. As discussed in Ref.~\onlinecite{Polson_2013}, the free energy
is characterized by an oscillation with an amplitude, $\Delta F_{\rm osc}$, that increases
with decreasing $R$. This is illustrated in Fig.~\ref{fig:F.N20.2R}(a),
which shows profiles for a $N=20$ polymer in a pore of length $L=6$ for 
pore radii of $R=0.55$ and $R=0.8$.
The origin of the oscillations is related to the nature of the variation with $Q$ of the 
orientational entropy of the bonds at the two edges of the nanopore for this model.
When the oscillation amplitude is of order $k_{\rm B}T$ or larger,
the translocation rate will be appreciably reduced. Thus, decreasing $R$ increases
$\Delta F_{\rm osc}$, which in turn increases $\langle\tau\rangle$. 
In addition to increasing the translocation times, the free energy oscillations can
have other effects on the translocation dynamics. Figure~\ref{fig:F.N20.2R}(b)
shows the history of $Q$ for a single translocation event for a $N=20$ polymer
in a pore of length $L=6$ and radius $R=0.55$. The polymer clearly undergoes
a diffusive type of motion while it is in the plateau region of the free energy
profile in Fig.~\ref{fig:F.N20.2R}.  Note, however, that the polymer tends
to dwell longer at $Q$ values close to the local free energy minima
and jump rapidly over the free energy barriers.
As $R$ decreases, the oscillation amplitude increases, and the increase
in the roughness of the landscape decreases the translocation rate.

\begin{figure}[!ht]
\includegraphics[width=0.40\textwidth]{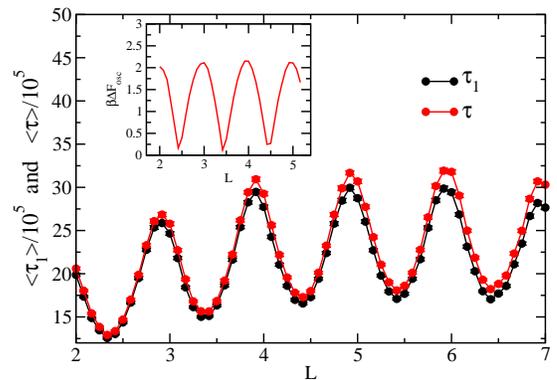}
\caption{Mean translocation times, $\langle\tau_1\rangle$ and 
$\langle\tau_\rangle$, vs nanopore length, $L$, for a polymer of length $N=20$ 
and a nanopore of radius $R=0.55$.  The inset shows the variation of the free 
energy oscillation amplitude, $\beta \Delta F_{\rm osc}$ vs $L$.}
\label{fig:tau.N=20.R=0.55}
\end{figure}

In addition to its dependence on pore radius, the amplitude of the free energy oscillations
also displays an interesting dependence on nanopore length. As is evident in the inset
of Fig.~\ref{fig:tau.N=20.R=0.55}, the amplitude itself oscillates with $L$ and has a period
of $\Delta L = 1$, i.e. one bond length. The origin of this effect was explained
in Ref.~\onlinecite{Polson_2013}. For the case of $R=0.55$ illustrated in the figure, 
the oscillation amplitude varies from negligible magnitude to a maximum of 
$\beta\Delta F_{\rm osc} \approx 2.1$. The effect on the variation of $\langle\tau_1\rangle$
and $\langle\tau\rangle$ with $L$ is shown in the figure and follows the expected pattern:
for high values of $\Delta F_{\rm osc}$, translocation proceeds more slowly and
the translocation times are high, while for nanopore lengths corresponding to 
low $\Delta F_{\rm osc}$, the translocation times are also low. 
This behaviour is expected to be independent of $N$, as long as the pore
friction is sufficiently high. Results for $N=100$ with $L=2.917$ and $L=2.333$
are consistent with this pattern (data not shown).

As explained in Section~\ref{sec:theory}, application of the FP formalism requires 
the value of the translocation diffusion coefficient or, equivalently, the
friction coefficient, $\gamma$, to predict translocation times. Since the friction 
coefficient cannot easily be determined, we employ the procedure described earlier.
The value of $\gamma$ is chosen such that the mean translocation 
time, $\langle\tau_1\rangle$, calculated using Eq.~(\ref{eq:tauQ0}) is equal to
the value obtained directly from the dynamics simulations.
This value can then be used in Eq.~(\ref{eq:gN2Np}) to calculate $N_{\rm eff}$,
whose value should be close to the estimate in Eq.~(\ref{eq:Npapprox}). Deviations
from this estimate can be interpreted as a breakdown in the quasi-static condition
and/or the assumption that pore friction is the dominant contribution to the total friction.
The scaling $\gamma\propto N^2$ predicted by Eq.~(\ref{eq:gN2Np}) provides another 
such test of quasi-equilibrium. Finally, this value of $\gamma$ obtained by
fixing $\langle\tau_1\rangle$ can be used in Eqs.~(\ref{eq:FP}) and (\ref{eq:gQ0tau})
to predict the complete distribution of times, $g(\tau;Q_0)$.

\begin{figure}[!ht]
\includegraphics[width=0.40\textwidth]{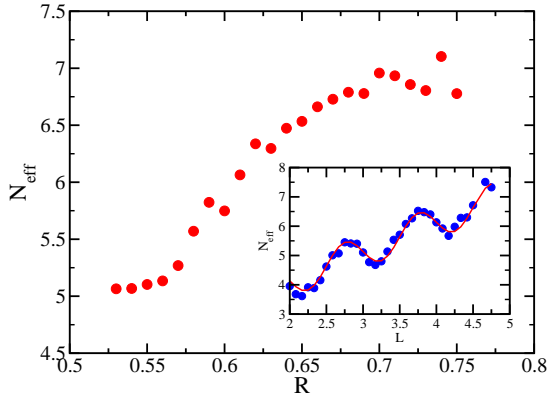}
\caption{Effective number of monomers in the nanopore, $N_{\rm eff}$, vs pore
radius for a polymer of length $N=20$ in a nanopore of length $L=3$. The inset
shows the variation of $N_{\rm eff}$ with $L$ for the same $N=20$ polymer in
a pore of radius $R=0.55$. The red curve is a fit to Eq.~(\ref{eq:sinfit}).}
\label{fig:Neff_N20_L3}
\end{figure}

Figure~\ref{fig:Neff_N20_L3} shows the variation of the calculated $N_{\rm eff}$ with
$R$ for a $N=20$ polymer in a pore of length $L=3$. At the high end of the range 
of values of $R$ shown, $N_{\rm eff}$ is significantly larger than the value
of $N_{\rm eff}\approx 5$ predicted by Eq.~(\ref{eq:Npapprox}). The 
average number of monomers inside the pore was observed to increase only slightly
with $R$, and so it does not account account for this result. Consequently,
it must be due to decrease in $\gamma_0$ with $R$, illustrated in Fig.~\ref{fig:gamma.R}.
Generally, $N_{\rm eff}$
decreases with decreasing $R$ for $R>0.56$. Below this value, it is approximately
constant, with a value $N_{\rm eff} \approx 5.1$. This suggests that the quasi-static 
condition holds in this low-$R$ regime. This interpretation is consistent with
the variation of the scaling exponent $\alpha$ with $R$ in Fig.~\ref{fig:tau.Nscale.L=2.917},
where the quasi-static limit value of $\alpha=2$ is observed for $R=0.55$.
The inset shows the variation of $N_{\rm eff}$
with $L$ for the same $N=20$ polymer and a pore of radius $R=0.55$. 
While $N_{\rm eff}$ does tend to increase with $L$, 
there is a superimposed oscillation with a period of $\Delta L=1$. 
To help understand these results, the variation of $N_{\rm eff}$ with $L$ was fit 
to the function 
\begin{eqnarray}
N_{\rm eff} = L/\sigma + a_0 - a_1 \sin(2\pi L/\sigma).
\label{eq:sinfit}
\end{eqnarray}
This function provides a reasonable fit over the range of $L$ considered, 
which yielded values of $a_0=2.18$ and $a_1=0.564$. Ignoring the oscillation, 
the prediction of $N_{\rm eff} = L/\sigma + 2.18$ is roughly consistent with 
the prediction of Eq.~(\ref{eq:Npapprox}). 
%
%
The origin of the oscillation is not completely understood at present but it
may be a result of the MC algorithm employed. Note
that the values of $a_0$ and $a_1$ are such that $N_{\rm eff}\geq L/\sigma+1.62$.
The lower limit exceeds the average number of monomers inside the nanopore.
Thus, the oscillation in $N_{\rm eff}$ is
associated with the reduced mobility of monomers that lie outside the pore, the
most significant of which lie in the zones of partial confinement near the pore
edges ($-\sigma/2<z<0$ and $L < z < L+\sigma/2$). This effect is similar to that of
the periodic variation of $\Delta F_{\rm osc}$ with $L$ illustrated in 
Fig.~\ref{fig:tau.N=20.R=0.55}, which is also associated with the monomers near 
the edges.\cite{Polson_2013}

\begin{figure}[!ht]
\includegraphics[width=0.40\textwidth]{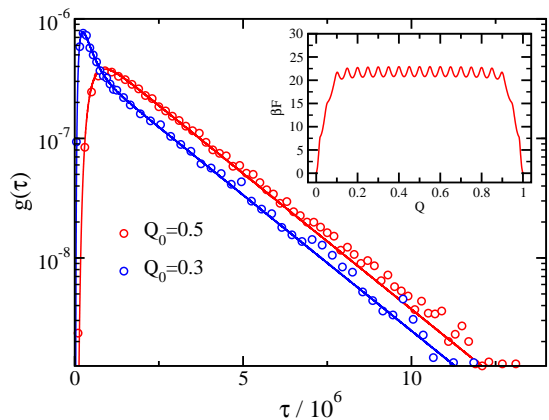}
\caption{Translocation time distributions for a polymer of length $N=20$
and a nanopore of radius $R=0.55$ and length $L=3$. Distributions for two
different $Q_0$ values are shown. The solid curves are theoretical predictions
using $N_{\rm eff}=5.1$, which is obtained by setting the mean predicted
time to be equal to the mean time from the dynamics simulations.
The inset shows the free energy function for this system.}
\label{fig:taudist_N20_R0.55_Q}
\end{figure}

Figure~\ref{fig:taudist_N20_R0.55_Q} shows translocation distributions
calculated for a $N=20$ polymer and a nanopore of dimensions $L=3$ and $R=0.55$.
Distributions for $Q_0=0.3$ and $Q_0=0.5$ are shown.  
The figure also shows theoretical predictions for the distributions, calculated
using the method described in Sections~\ref{sec:theory} and \ref{subsec:theory-calculations}. 
The friction coefficient $\gamma$ was determined using Eq.~(\ref{eq:gN2Np}) with
$N_{\rm eff}=5.1$, the value obtained by constraining the predicted $\langle\tau_1\rangle$ 
to be equal to that measured directly from the MC dynamics simulations for
$Q_0=0.5$. The theoretical distributions are in excellent quantitative agreement with 
those obtained by simulation.  

The translocation time distributions in Fig.~\ref{fig:taudist_N20_R0.55_Q} each
display exponential tails at longer times. The origin of this feature 
can be understood in the context of simple diffusion between
absorbing boundaries. For unforced translocation, the polymer spends most of its time 
moving over a relatively flat plateau of the free energy function, which is shown in 
the inset of the figure. When it reaches the plateau edge, the polymer
quickly exits the pore during the pore-emptying stage, i.e, 
$\tau=\tau_1+\tau_2\approx \tau_1$, as described above.
Neglecting the oscillations and the very slight overall negative curvature of 
the plateau, this process can be viewed approximately as simple diffusion with absorbing
boundary conditions at $Q_a=0.1$ and $Q_b=0.9$. In this case, Eqs.~(\ref{eq:FP}) and
(\ref{eq:gQ0tau}) can be solved analytically, yielding:\cite{Muthukumar_book}
\begin{eqnarray}
g(\tau,Q_0) = \frac{2D_{\rm eff}}{L_Q} 
\sum_{p=1}^\infty && \beta_p \left[1-\cos(\beta_pL_Q)\right]
\sin(\beta_p (Q_0-Q_a)) \exp\left[-\beta_p^2D_{\rm eff}\tau\right],
\end{eqnarray}
where $\beta_p \equiv \pi p/L_Q$, and $L_Q\equiv Q_{\rm a} - Q_{\rm b}=0.8$.
For long first-passage time $\tau$, the $p=1$ term becomes dominant; consequently,
\begin{eqnarray}
g(\tau;Q_0) = C(Q_0) \exp(-\tau/\tau_{\rm c}),
\end{eqnarray}
where 
\begin{eqnarray}
C(Q_0) \equiv \frac{4\pi D_{\rm eff}}{L_Q^2} \sin(\pi ((Q_0-Q_{\rm a})/L_Q),
\end{eqnarray}
and where the time constant, $\tau_c$, is given by
\begin{eqnarray}
\tau_{\rm c} = L_Q^2/(\pi^2D_{\rm eff}).
\label{eq:taucLQ}
\end{eqnarray}
Thus, the exponential decay time is predicted to be independent of $Q_0$, which
is consistent with the data in the figure. Fitting the data in the exponential region to
the function 
\begin{eqnarray}
g(\tau) = C' \exp(-\tau/{\tau_{\rm c}'})
\label{eq:expfit}
\end{eqnarray}
yields $\tau_{\rm c}'\approx 1.98\times 10^6$. Using this value in Eq.~(\ref{eq:taucLQ}),
we estimate $D_{\rm eff}\approx 3.2\times 10^{-8}$. Of course, this approximation
is strictly valid for a perfectly flat free energy curve. The corrugated
structure of the free energy plateau slows down the rate of diffusion, 
and thus this value is expected to be lower than the value of $D$ calculated using the 
methods described in Section~\ref{sec:theory}. Using Eqs.~(\ref{eq:Dg}) and 
(\ref{eq:gN2Np}) with a value $N_{\rm eff}=5.1$ yields a value of $D=4.84\times 10^{-8}$,
which is greater than $D_{\rm eff}$, as expected.
The fit also yields $C'=6.7\times 10^{-7}$ for $Q_0=0.5$, and $C'=4.5\times 10^{-7}$
for $Q_0=0.3$. These values are approximately consistent with the values predicted using 
Eq.~(\ref{eq:taucLQ}) of $C(Q_0=0.5) = 6.3\times 10^{-7}$ and $C(Q_0=0.3)=4.4\times 10^{-7}$.
The small discrepancies arise from the approximations employed.

For the distributions shown in Fig.~\ref{fig:taudist_N20_R0.55_Q}, it is evident that the mean
translocation time for $Q_0=0.3$ is lower than the mean time for $Q_0=0.5$.
This is an illustration of a general result: the closer the polymer is initially situated 
to either one of the pore exits, the shorter the translocation time.
This arises because the polymer exits more frequently
at the closer pore boundary, and the shorter distance reduces the first passage time.
The longer time required to reach the more distant exit is offset
by the diminished probability of reaching that exit. This is consistent 
with the predictions for simple diffusion between
absorbing boundaries.\cite{Muthukumar_book} It also effects a shift 
of the entire distribution to shorter times. 

\begin{figure}[!ht]
\includegraphics[width=0.40\textwidth]{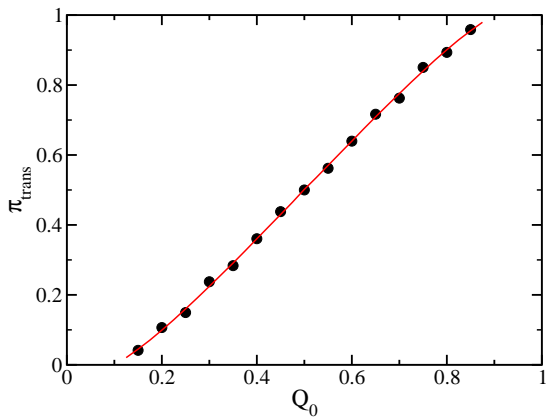}
\caption{Translocation probability, $\pi_{\rm trans}$ vs initial coordinate $Q_0$ for
a polymer of length $N=20$ and a nanopore of radius $R=0.55$ and length $L=3$.
The symbols are data calculated directly from the MC dynamics simulations,
and the solid line is a theoretical prediction using Eq.~(\ref{eq:pib}).}
\label{fig:pab.N=20.L=3.R=0.55}
\end{figure}

Figure.~\ref{fig:pab.N=20.L=3.R=0.55} shows the variation with $Q_0$ of the 
probability of a {\it trans} side exit.
As the initial position is moved closer to the {\it cis} side, i.e. as $Q_0$
decreases, the probability of a {\it trans} exit decreases, and the {\it cis}
exit probability increases accordingly.  
The theoretical curve was calculated using Eq.~(\ref{eq:pib}), with $Q_{\rm a}=0.1$
and $Q_{\rm b}=0.9$, the bounds of the free energy plateau shown in the inset
of Fig.~\ref{fig:taudist_N20_R0.55_Q}. The theoretical prediction is in excellent
agreement with the simulation probabilities. 
The observed variation of $\pi_{\rm trans}$ with initial position corresponds closely
to the prediction for a diffusive process with a constant free energy
between two absorbing boundaries, where the variation is exactly 
linear.\cite{Muthukumar_book}
The slight deviation from linearity observed in Fig.~\ref{fig:pab.N=20.L=3.R=0.55} arises
from the overall slight negative curvature of the (rough) free energy landscape 
evident in the inset in Fig.~\ref{fig:taudist_N20_R0.55_Q}.

\begin{figure}[!ht]
\includegraphics[width=0.40\textwidth]{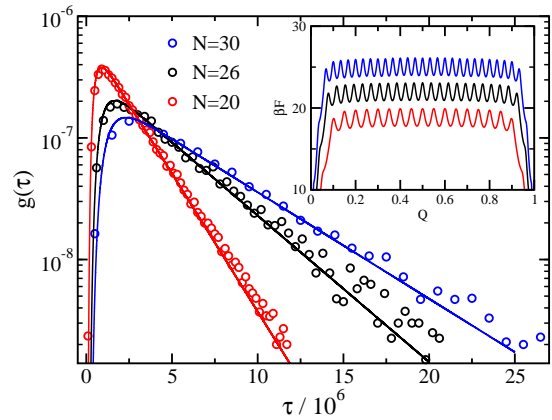}
\caption{Translocation time distributions for polymers of lengths $N=20$, 26
and 30, and a nanopore of radius $R=0.55$ and length $L=3$. 
The solid curves are theoretical predictions using $N_{\rm eff}=5.1$.
The inset shows the free energy functions for the three systems. The functions
have been vertically shifted to prevent overlap for visual clarity.}
\label{fig:taudist_N_R0.55}
\end{figure}

Figure~\ref{fig:taudist_N_R0.55} shows translocation time distributions 
calculated for three different polymer lengths using a nanopore with $L=3$ and $R=0.55$
and an initial coordinate of $Q_0=0.5$. As the polymer length increases, the distribution 
tends to broaden and the mean value increases.  The theoretical distributions overlaid
on the simulation data were all calculated using a friction coefficient, $\gamma$, 
obtained using Eq.~(\ref{eq:gN2Np}) with $N_{\rm eff}=5.1$. 
As explained in Section~\ref{sec:theory}, the value of $N_{\rm eff}$ is
expected to depend on the nanopore dimensions, but not on the polymer length. The 
excellent agreement between the theoretical distributions and those calculated by MC dynamics
simulations using the same $N_{\rm eff}$ is consistent with this expectation.

\begin{figure}[!ht]
\includegraphics[width=0.40\textwidth]{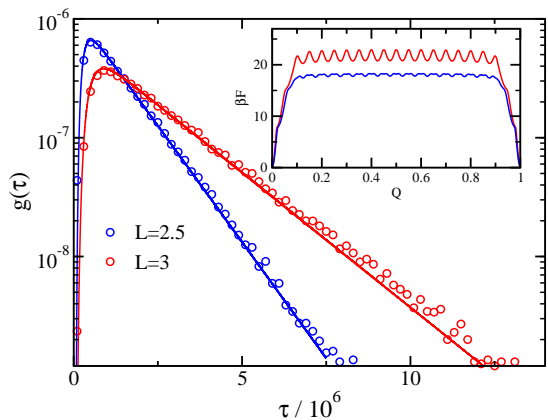}
\caption{Translocation time distributions for a nanopore of radius $R=0.55$ and 
length $L=3$. The polymer started midway through the pore with $Q_0=0.5$.
Distributions for two different nanopore lengths are shown.
The solid curves are theoretical predictions 
using $N_{\rm eff}=5.1$ for $L=3.0$ and $N_{\rm eff}=4.6$
for $L=2.5$.} 
\label{fig:taudist_N20_R0.55_Rclarge}
\end{figure}

Figure~\ref{fig:taudist_N20_R0.55_Rclarge} shows a comparison of translocation
distribution times for nanopore lengths of $L=3$ and $L=2.5$, for the case where
the polymer length is $N=20$ and the nanopore radius is $R=0.55$. The distribution
is shifted to lower times for the shorter nanopore. The theoretical curves shown
in the figure show excellent agreement with the simulation data.
Note that the theoretical curve for $L=2.5$ was calculated using $N_{\rm eff}=4.6$, 
a value that is consistent 
with the estimate in Eq.~(\ref{eq:Npapprox}). A fit of the exponential tail of the 
$L=2.5$ distribution yielded a time constant of $\tau_{\rm c}'=1.135\times 10^{6}$.
As seen in the figure inset, the free energy function is much smoother over the 
plateau region for $L=2.5$ than for $L=3$. In addition, the width of the plateau
is approximately the same, i.e. $L_{\rm Q} \approx 0.8$ in each case. Employing
the model of free diffusion between absorbing boundaries at $Q=0.1$ and 0.9
for the case of $L=2.5$, Eq.~(\ref{eq:taucLQ}) yields a value of 
$D_{\rm eff}=5.7\times 10^{-8}$, which is very close to the value $D=5.48\times 10^{-8}$
predicted using Eqs.~(\ref{eq:Dg}) and (\ref{eq:gN2Np}). This is exactly the
consistency expected for a smooth, flat free energy function, and stands in contrast
to the discrepancy between similarly calculated values of $D_{\rm eff}$ and $D$ for 
$L=3$ discussed above in the context of Fig.~\ref{fig:taudist_N20_R0.55_Q}. It follows that
much of the reduction in the translocation time upon decreasing the nanopore length
from $L=3$ to $L=2.5$ is accounted for by decreasing roughness of the free energy 
function; a smaller contribution arises from the small decrease in $N_{\rm eff}$
and corresponding small increase in $D$.

\begin{figure}[!ht]
\includegraphics[width=0.40\textwidth]{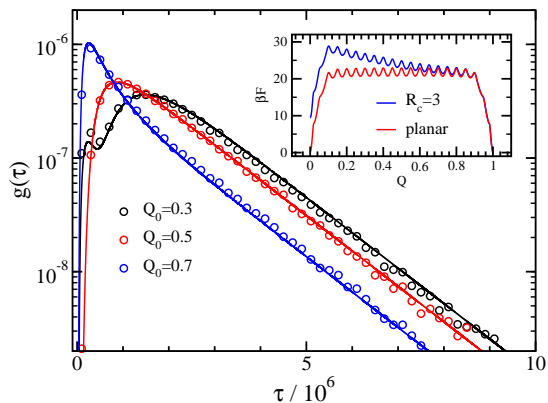}
\caption{Translocation time distributions for the spherical cavity
system as illustrated in Fig.~\ref{fig:geometry}(b) for a cavity radius
of $R_{\rm c}=3$.  In addition, $R=0.55$, $L=3$ and $N=20$.
The theoretical curves were calculated as described in the text, using $N_{\rm eff}=5.6$. 
The inset shows the free energy function used in these theoretical calculations
as well that used in calculations for the planar geometry.}
\label{fig:taudist.Rc}
\end{figure}

Figure~\ref{fig:taudist.Rc} shows translocation time distributions for a polymer
that translocates out of a spherical cavity, the system illustrated in 
Fig.~\ref{fig:geometry}(b).  The cavity radius is $R_{\rm c}=3$, and as before, $N=20$,
$L=3$, and $R=0.55$. Distributions for three different
$Q_0$ are shown. Comparing the results with those of Fig.~\ref{fig:taudist_N20_R0.55_Q} 
for the flat barrier system shows that translocation times are slightly lower
in the spherical cavity system.  The origin of this effect is clear from
a comparision of the free energy functions, shown in the figure inset.
The confinement of the polymer in the cavity breaks the symmetry and leads to a 
free energy gradient. The resulting entropic driving force pushes the polymer
to the {\it trans} side and reduces the time taken for the polymer to exit the nanopore. 

The asymmetry in the free energy function means that $Q_0=0.3$ and $Q_0=0.7$ 
no longer have the same distributions, unlike 
the case for the symmetric flat-barrier system of Fig.~\ref{fig:geometry}(a).
Instead, there is a monotonic decrease in the distribution mean times
as the starting point of the polymer is placed closer to the {\it trans} side
of the nanopore (i.e. larger $Q_0$). The bias introduced by the entropic driving
force causes the majority of translocation events associated with a polymer exit
on the {\it trans} side of the pore. The closer to the {\it trans} side
the polymer starts, the faster will be its exit. Note that this trend must inevitably
reverse for sufficiently low $Q_0$, i.e. if the polymer is almost complete inside
the {\it cis} cavity, in which case the polymer may exit very rapidly into the spherical
cavity.  

The theoretical distributions shown in Fig.~\ref{fig:taudist.Rc}
are in excellent agreement with the simulation data.  Note that the theoretical
distribution for the spherical cavity system shown in the figure was calculated using
$N_{\rm eff}=5.6$, somewhat larger than the value of $N_{\rm eff}=5.1$ employed for the
flat-barrier system. As discussed above, $N_{\rm eff}$ is expected to depend only on
$L$, which has the same value ($L=3$) as that used in the flat barrier simulations. 
The slightly higher value of $N_{\rm eff}$ could be a result of out-of-equilibrium 
effects which arise from more rapid translocation due to the entropic driving
force, as well as an increase in the relaxation time of the section of the 
polymer inside the spherical cavity.  If so, then $D$ and $\gamma$ would likely
exhibit a dependence on polymer length. As a check, we calculated the
translocation time distributions for a longer polymer, $N=30$, for an 
otherwise identical system. The theoretical predictions are in comparably
good quantitative agreement with the distributions calculated from the
dynamics simulations for the same value of $N_{\rm eff}$ (data not shown).
This suggests that the difference in the values of $N_{\rm eff}$ 
for the two systems is not an out-of-equilibrium effect. More likely, 
the effect arises from the curvature
of the cavity wall in the vicinity of the {\it cis} entrance to the nanopore.
The curvature should lead to a slightly more crowded region near the entrance,
which will lead to a slightly lower acceptance rate in the MC algorithm due to
an increased likelihood of particle overlap. This in turn leads to a decrease in 
the monomer mobility near the entrance and consequently a slightly larger $\gamma$
and $N_{\rm eff}$. It has already been noted in the discussion above regarding
the data in Fig.~\ref{fig:Neff_N20_L3} how sensitive the pore friction
can be to very small changes in the nanopore characteristics.

\begin{figure}[!ht]
\includegraphics[width=0.40\textwidth]{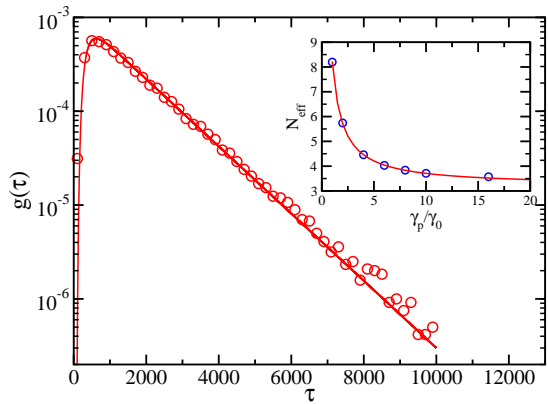}
\caption{Translocation time distribution for a repulsive LJ chain 
for the nanopore geometry of Fig.~\ref{fig:geometry}(a) calculated
using Brownian dynamics simulations. Here, $\gamma_{\rm p}/\gamma_{\rm 0}=16$,
$R=0.65$, $L=3$, and $N=20$. The solid line is the theoretical curve 
calculated using the FP formalism.  The inset shows the variation of 
$N_{\rm eff}$ with $\gamma_{\rm p}/\gamma_0$.}
\label{fig:taudist.lj}
\end{figure}

The FP predictions of translocation time distributions are in 
excellent quantitative agreement with those measured using MC dynamics
method for the hard-potential model described in Section~\ref{sec:model}.
In order to demonstrate the generality of this result, we present
preliminary results for a repulsive Lennard-Jones chain model.
The details of the model are essentially the same as in Ref.~\onlinecite{deHaan_2010},
except that nanopore of finite length is used. The free energy function 
used in the FP calculation was calculated as in Ref.~\onlinecite{Polson_2013}, 
and Brownian dynamics simulations were used to measure the distributions.
Quasi-static conditions were achieved by increasing the friction coefficient
of monomers inside the pore, $\gamma_{\rm p}$, relative to that for 
monomers ouside, $\gamma_0$. Figure~\ref{fig:taudist.lj} shows results
for $\gamma_{\rm p}/\gamma_{\rm 0}=16$, for which there is excellent
quantitative agreement between theory and simulation. The inset
shows that the measured $N_{\rm eff}$ approaches the expected value
for sufficiently large $\gamma_{\rm p}/\gamma_0$. The results demonstrate
that the FP methodology provides a valid description of quasi-static
translocation dynamics for this commonly used simulation model.
Further details will be presented in a future study.

\section{Discussion}
\label{sec:discussion}

In this work, we have studied the translocation dynamics of a simple 
polymer/nanopore model system using Monte Carlo dynamics simulations 
and the Fokker-Planck formalism. The emphasis was on the quasi-static regime, 
in which the FP formalism provides a valid approach for describing translocation.
While many theoretical studies have employed this approach, 
almost all have used approximate analytical forms for the free energy 
function. In our study, we use free energy functions calculated directly using a 
MC multiple histogram method for the same model used in the dynamics simulations.
The translocation time distributions calculated using these functions thus
provide a much better test of this theoretical approach than would otherwise
be possible.

For the MC dynamics algorithm employed here, the key parameter governing the
dynamical regime is the nanopore radius, $R$. As noted in a previous 
study,\cite{deHaan_2010}
decreasing $R$ can slow the translocation rate in a manner that leads to a
decrease in the translocation time scaling exponent, $\alpha$. In the present case,
this happens in part by effectively increasing the friction of the polymer in the pore.
For sufficiently low $R$, the translocation rate decreased to the point that a value of 
$\alpha=2$ was observed, indicating that the quasi-static regime was attained. 
This effect is similar to that observed in Ref.~\onlinecite{deHaan_2012b},
in which decreasing the solvent viscosity was used to attain 
the same exponent value. In each case, quasi-equilibrium is observed when the
ratio of the translocation time to the conformational relaxation times of the 
{\it cis} and {\it trans} chain sections is sufficiently high. Muthukumar 
has noted that this is the relevant dynamical regime for experiments of DNA 
translocation through $\alpha$-hemolysin pores.\cite{Muthukumar_book}

Knowledge of the translocation free energy functions can provide valuable insight into
the quantitative trends of the translocation dynamics. An useful illustration
in this study is the effect of oscillations in the functions that occur for
sufficiently small pore radius, $R$.\cite{Polson_2013} 
The translocation time is generally longer for systems with a higher free energy 
oscillation amplitude. Interestingly, the oscillation amplitude itself can be very
sensitive to the nanopore length, $L$, as is evident in Fig.~\ref{fig:tau.N=20.R=0.55},
which in turn leads to a sensitive dependence of the translocation time on $L$.
While it is unlikely that oscillations arising from the same physical origin
are present in the systems studied experimentally\cite{Oscillations_note} this
effect may be highly relevant to interpretation of results from other simulation 
studies, many of which use qualitatively similar models to that employed here.
Since the oscillations are associated with the orientational freedom of 
only two polymer segments localized at the nanopore edges, the effect of slowing
the translocation rate may even be present in cases where the {\it trans} and/or
{\it cis} chain sections are out of equilibrium. Consequently, small changes in
the nanopore geometry (such as $L$, here) could lead to significant changes in
the translocation times, with corresponding changes in the scaling exponent,
even outside the quasi-static regime. Such effects may account in part for the
wide range of scaling exponents reported in various simulation studies.
This underscores the importance of calculating accurate free energy functions 
to aid in the interpretation of dynamics simulation data.

Translocation time distributions for systems determined to be in the quasi-static regime
were calculated using the FP formalism with measured free energy functions.  
The calculations assume that the nanopore friction is the dominant contribution 
to the overall friction coefficient for the translocation process.  The distributions 
were calculated using a coefficient $\gamma$ which was characterized 
by the parameter $N_{\rm eff}$, the effective number of monomers whose motion is 
strongly affected by nanopore confinement.  In the case
of unforced translocation, we investigated the effects of varying polymer length, $N$, 
nanopore length, $L$, and the initial position of the polymer, $Q_0$. In addition, we 
measured distributions for polymers translocating out of a spherically confined space, a 
feature that gives rise to an entropic force that drives translocation. The calculated 
distributions were generally in excellent quantitative agreement with those obtained 
directly from dynamics simulations for physically meaningful values of $N_{\rm eff}$.
Preliminary results of calculations using a repulsive LJ chain chain model 
and Brownian dynamics simulations yielded comparably good agreement.

In this study, calculations were carried out mainly for short polymer chains. This
was due to the considerable computational effort required to obtain very accurate 
free energy functions and translocation time distributions from the simulations. 
In future work, we will investigate the effectiveness of FP formalism in describing
translocation dynamics of much longer polymers. Another important issue not adequately
addressed in this study is the effect of a driving force, which plays an essential role in
DNA translocation experiments and whose effects have been extensively investigated
in other simulation studies. Though we have briefly considered the effect of an 
entropic driving force in this work, a driving force associated with an potential 
energy gradient across the nanopore is not suitable for the MC dynamics algorithm employed 
here.\cite{Driving_force_note} In a future study, we will examine this effect using Langevin 
and Brownian dynamics simulations.  In addition, we will more thoroughly
elucidate the regime within which the quasi-static approximation is sufficiently valid
to justify using the FP formalism. This regime is expected to show a complex
dependence on the polymer length, the conformational relaxation times of the chain, 
pore friction, the driving force strength, as well subtle features of 
the free energy functions. This work will also be extended to using more realistic 
molecular models to provide a more meaningful comparison with experimental systems.
The long term goal of this research is to provide accurate predictions of translocation
times using precisely calculated free energy functions within a well defined regime
of validity.

\begin{acknowledgments}
J.M.P. would like to thank Sheldon B. Opps for helpful discussions.
This work was supported by the National Research Council of Canada (NSERC).
We are grateful to the Atlantic Computational Excellence Network (ACEnet) for use of 
their computational facilities.
\end{acknowledgments}

%
%


\begin{thebibliography}{88}
\bibitem [{Dri()}]{Driving_force_note}%
  \BibitemOpen
  \href@noop {} {}\bibinfo {note} {Consider the drift-diffusion process for a
  polymer in an infinite tube with a constant energy gradient. The polymer
  driving force is $\propto N$. On the other hand, from Eq.~(\ref{eq:gz0_inf}),
  the friction coefficient associated with diffusion is $\gamma_z = D_z/k_{\rm
  B}T \propto N-1$. This leads to an inconsistency between the friction
  coefficient measured from diffusion and that measured from the polymer drift
  velocity. This is also true for tubes of finite length. In the Langevin
  dynamics algorithm, $\gamma_z = D_z/k_{\rm B}T \propto N$, and this problem
  does not arise.}\BibitemShut {Stop}%
\end{thebibliography}

%

\end{document}